# A Continuum Poisson-Boltzmann Model for Membrane Channel Proteins

Li Xiao[1], Jianxiong Diao[2], D'Artagnan Greene[2],
Junmei Wang[3], and Ray Luo[1,2,4,5]

1. Department of Biomedical Engineering, 2. Department of Molecular Biology and Biochemistry, 4. Chemical and Materials Physics Graduate Program,
5. Department of Chemical Engineering and Materials Science,
University of California, Irvine, CA 92697,
3. Department of Pharmaceutical Sciences, University of Pittsburg, Pittsburgh, PA 15261

Membrane proteins constitute a large portion of the human proteome and perform a variety of important functions as membrane receptors, transport proteins, enzymes, signaling proteins, and more. The computational studies of membrane proteins are usually much more complicated than those of globular proteins. Here we propose a new continuum model for Poisson-Boltzmann calculations of membrane channel proteins. Major improvements over the existing continuum slab model are as follows: 1) The location and thickness of the slab model are fine-tuned based on explicit-solvent MD simulations. 2) The highly different accessibility in the membrane and water regions are addressed with a two-step, two-probe grid labeling procedure, and 3) The water pores/channels are automatically identified. The new continuum membrane model is optimized (by adjusting the membrane probe, as well as the slab thickness and center) to best reproduce the distributions of buried water molecules in the membrane region as sampled in explicit water simulations. Our optimization also shows that the widely adopted water probe of 1.4 Å for globular proteins is a very reasonable default value for membrane protein simulations. It gives an overall minimum number of inconsistencies between the continuum and explicit representations of water distributions in membrane channel proteins, at least in the water accessible pore/channel regions that we focus on. Finally, we validate the new membrane model by carrying out binding affinity calculations for a potassium channel, and we observe a good agreement with experiment results.

Please send correspondence to: ray.luo@uci.edu

# Introduction

Membrane proteins constitute a large portion of the human proteome and perform a variety of important functions, such as membrane receptors, transport proteins, enzymes, and signaling proteins [1]. These important proteins have become primary drug targets in modern medicine: over 60% of all drugs target these proteins [2-4]. However, the study of membrane proteins is usually much more complicated than that of globular proteins, both experimentally and computationally. For experimental studies, the difficulty of obtaining a high-resolution structure is an obstacle, especially for studies that involve proteins found in humans. For computational studies, modeling of the membrane environment is also an important consideration.

Since most biomolecular systems exist in an aqueous environment, it is important to account for solvent effects. There are two ways to include solvent effects in a computational simulation: explicit and implicit solvation. In explicit solvation modeling, each solvent atom is modeled explicitly. Although this is the most accurate method, what we are interested in is often not the properties of the solvent itself, but rather its influence on the solute molecules. In addition, accurately capturing the solvent influence in a statistically meaningful way requires sampling either from an ensemble of trajectories or from a single very long trajectory, which is very computationally demanding. Implicit solvation modeling provides an attractive alternative wherein the solvent molecules are collectively modeled as a continuum. In implicit solvent models, although the details of individual solvent atoms are lost, the relevant important statistically averaged effects can still be preserved by design. Since solvent molecules typically constitute the major portion of molecules for an explicit solvent simulation, implicit solvent modeling can lead to much more efficient simulations [5-21]. In addition to water, membrane molecules should also be included when modeling solvation effects, and implicit membrane modeling has also been developed [22-28].

A key issue in developing implicit solvent models is the modeling of electrostatic interactions. The Poisson-Boltzmann equation (PBE) has been established as a fundamental equation to model continuum electrostatic interactions [29-47]. The solvent molecules are modeled as a continuum with a high dielectric constant, and the solute atoms are modeled as a continuum with a low dielectric constant and buried atomic charges. The effect of charged ions in the solvent region is included by adding mobile charge density terms that obey Boltzmann distributions. The potential of the full system is then governed by the partial differential equation:

$$\nabla \cdot \varepsilon \nabla \phi = -4\pi\rho_0 - 4\pi \sum_i ez_i c_i \lambda \exp(-ez_i\phi/k_B T) \qquad (1)$$

where $\nabla$ is the spatial gradient operator, $\varepsilon$ is the dielectric constant distribution, $\phi$ is the electrostatic potential distribution, $\rho_0$ is the charge density of the solute (usually modeled as a set of discrete point charges), $c_i$ is the concentration of the $i$th solvent ion species in bulk, $e$ is the absolute charge of an electron, $z_i$ is the valence for the $i$th ion, $k_B$ is Boltzmann's constant, $T$ is the temperature, and $\lambda$ is the Stern layer masking function, which is 0 within or 1 outside of the Stern layer.

The PBE is a non-linear elliptical partial differential equation. There is no closed form solution, and thus, numerical methods are often required for biomolecular applications [22, 36, 44, 45, 48-87]. Efficient numerical PBE-based solvent models have been widely used to study biological processes including predicting p$K$a values [88-91], computing solvation and binding free energies [92-101], and protein folding [102-112]. Predicting protein-ligand binding affinities is one of the major applications for implicit solvent free energy calculations. In the Amber software package, MMPBSA is the module performing such calculations [113-118]. Implicit membrane modeling has also been applied and developed in binding free energy calculations. There are a noticeable number of pioneer works that implement implicit membrane modeling in several PB packages, such as APBS [27], Delphi [28, 119], PBEQ [78, 120], and PBSA [121-123]. All of them add the membrane as a slab with a relatively low dielectric constant that is embedded in water for PBE calculations.

Our previous work implemented an implicit membrane model into the PBE framework [121]. The implicit membrane model can be readily interfaced with the existing MMPBSA program [113-118] to perform binding free energy calculations of several protein structures embedded in a membrane [87, 124]. However, a problem arises when those membrane proteins contain a pore or a gated channel, since the region of the channel is usually permeable and should be composed of water. Therefore, a simple slab-like membrane setup may cause problems if the membrane protein contains pore- or channel-like region(s). Similar to the approaches adopted in the community [28, 119, 125], we dealt with this issue by manually defining the pore region as a cylinder, and we then set the dielectric constant within the cylindrical region as that of water if it was not occupied by protein atoms. The limitation of this method is that, for every snapshot of a trajectory, we need to visualize and locate the cylinder by hand, which is neither efficient nor practical given the large number of snapshots that must be processed for converged calculations.

In this work, we propose a new continuum membrane model for PBE calculations of biomolecules. Major improvements from the existing continuum slab model are the following: 1) an explicit solvent MD simulation was exploited to fine tune the slab model, i.e. its exact location and thickness, to best reproduce the solvent accessibility and the water accessible channel, 2) a two-step, two-probe initial grid labeling procedure was adopted to address highly different accessibility in the membrane region and water region, and 3) a depth-first search algorithm was introduced to detect the water pores/channels automatically based on the initial grid labels. This procedure follows our basic algorithm proposed for globular proteins, and adds little overall overhead in the application of linear finite-difference PBE solvers to typical membrane proteins.

## Methods

The Poisson-Boltzmann equation (Eqn (1)) is widely used in capturing electrostatic energy and forces in implicit solvent modeling. For systems with dilute ion concentrations, the second term on the right-hand side is usually linearized, giving the simpler form:

$$\nabla \cdot \varepsilon \nabla \phi = -4\pi \rho_0 + \lambda \kappa^2 \phi \tag{2}$$

where $\kappa^2 = 4\pi \sum_i c_i e_i^2 z_i^2 / k_B T$.

The finite-difference method [22, 43, 71-83, 86] is one of the most popular methods used in the numerical implementation of the PBE. In a typical procedure, the rectangular grid covering the solution system is first defined. Next, the atomic point charges are mapped onto the grid points with a predefined assignment function. Third, the dielectric constant distribution is mapped to the grid edges. The discretized linear system is then turned to a linear solver to solve for potentials on the grid points, which can be expressed as:

$$\begin{bmatrix} \begin{bmatrix} \varepsilon_x(i,j,k)\phi(i+1,j,k)+\varepsilon_x(i-1,j,k)\phi(i-1,j,k) \\ \varepsilon_y(i,j,k)\phi(i,j+1,k)+\varepsilon_y(i,j-1,k)\phi(i,j-1,k) \\ \varepsilon_z(i,j,k)\phi(i,j,k+1)+\varepsilon_z(i,j,k-1)\phi(i,j,k-1) \end{bmatrix} - \\ \begin{bmatrix} \varepsilon_x(i-1,j,k)+\varepsilon_x(i,j,k)+\varepsilon_y(i,j-1,k) \\ \varepsilon_y(i,j,k)+\varepsilon_z(i,j,k-1)+\varepsilon_z(i,j,k) \end{bmatrix}\phi(i,j,k) \end{bmatrix} + h^2\lambda(i,j,k)\kappa^2\phi(i,j,k) = -\frac{4\pi\rho(i,j,k)}{h} \quad (3)$$

Here $\varepsilon_x, \varepsilon_y$, and $\varepsilon_z$ represent the dielectric constants for grid edges along the x, y, and z directions respectively, and *h* represents the grid spacing.

This study focuses on how to set up linear PBE applications for membrane systems. A major issue is the presence of the membrane and its influence on the dielectric constant distribution. In globular proteins, the solvent excluded surface (SES)[43, 126-130] is often used as a boundary separating the high dielectric water exterior and the low dielectric protein interior. The presence of the membrane introduces at least a third region. In this study, we adopt the uniform membrane dielectric model, though our procedure can be easily extended to accommodate another often used depth-dependent membrane dielectric model.

The first step is to introduce a membrane region to the existing solvent excluded surface procedure with minimum invasion to the program and minimum efficiency lost. The SES is the most common surface definition used to describe the dielectric interface between the two piece-wise dielectric constants. In fact, comparative analysis of PB-based solvent models and TIP3P solvent models have shown that the SES definition is reasonable in the calculation of reaction field energies and electrostatic potentials of mean force fields.[131-133] Here, we follow the idea from Rocchia *et al.*[130] and Wang *et al.*[43] of mapping the SES to a finite-difference grid. While keeping the variables used to label the solvent and solute regions, we also introduce a new variable to label the membrane region. Considering the membrane molecules are usually larger than solvent molecules, we use two different solvent probe radii to set up the membrane and solvent regions. And finally, we assign the dielectric constant on each region and the interface.

**Grid point labeling**

Our general strategy is to model the membrane as a second continuum solvent of finite region, i.e. a slab located at a user specified position. The essence of the algorithm is to determine both the membrane accessibility and water accessibility around a molecular solute. Assisted with both sets of accessibility data, the presence of water channels or water pores within the membrane region can then be identified in the next step. Due to the much larger size of lipid molecules, a separate solvent probe (**mprob**) must be used to determine the membrane accessibility. This is apparently much larger than the water probe (**dprob**). The influence of both probes on reproducing the solvent accessible surface of a membrane protein is presented in Results and Discussion.

In Amber/PBSA, an integer array **insas** is used to label whether the grid point is outside the solute region (**insas**<0) or inside the solute region (**insas**>0) for fast mapping of solvent accessibility information [43]. This labeling scheme has been extended to map all commonly used surfaces, SES, SAS, VDW, and DEN in recent Amber and AmberTools releases.[36, 43, 80, 82, 87, 134, 135] To minimize the interference to existing procedures and maximize efficiency, a separate integer array **inmem** is used to label whether a grid point is inside the membrane (**inmem**>0) or outside the membrane (**inmem**=0). Specifically, the grid labeling algorithm can be summarized as the following five steps:

0. Initialize **insas** of all grid points as "-4", *i.e.* in the bulk solvent and salt region, and **inmem** of all grid points as "0", *i.e.* outside the membrane region.
1. Using **mprob** as the solvent probe radius, label **insas** of all grid points as "-3" if within the stern layer; "-2" if within the solvent accessible surface layer; "-1" if within the reentry region but outside the SES; "1" if within the reentry region but inside the SES; "2" if inside the VDW surface.
2. Add a slab perpendicular to the z-axis as the membrane region at the specified location. Label **inmem** of the membrane-region grid points with **insas**<0 as "1".
3. Apply the depth-first search algorithm to detect any possible membrane accessible grid point that is not connected to the bulk membrane. If so, relabel its **inmem** as "0".
4. For each grid point with (**inmem**=0) within the slab, if it has a neighbor with (**inmem**=1) within the distance cutoff of **memmaxd**, relabel its **inmem** as "2".
5. Using **dprob** as the solvent probe radius, relabel **insas** of all grid points as "-3" if within the stern layer; "-2" if within the solvent accessible surface layer; "-1" if within the reentry region outside the SES; "1" if within the reentry region inside the SES; "2" if inside the VDW surface.

A few explanations are in order here. First, **inmem** is determined in Step 2 through Step 4, so that its value is controlled by both the **mprob**-generated **insas** and the depth-first search algorithm. Second, a new variable (**memmaxd**) is introduced in Step 4. Since **mprob** is usually much larger than **dprob**, there exists a thin layer of grid points with **insas**>0 and **inmem**=0 between the membrane region and the protein region. If the grid labels are set this way, these grid points would be labeled as water in a later processing stage of our method, thus leading to an artificial layer of water between the protein and membrane. To resolve this issue, a cutoff distance of **memmaxd** is introduced to represent the maximum difference between the SES surfaces generated by **mprob** and **dprob**. This is estimated to be **mprob**−**dprob** assuming maximum reentry by **dprob**. Thus Step 4 changes the **inmem** labels of the grid points from 0 (**mprob** inaccessible) to 2 (**mprob** accessible) if they are **memmaxd** inside the **mprob**-generated SES. The correction effectively removes the artificial layer of water between the protein and the membrane. Here the revised **inmem** values are set to be "2" so these grid points would not interfere with the subsequent search. Note too that this correction does not change the protein interior definition, which is defined with the water **dprob**. Nevertheless, it does have the effect of pushing back the potential buried water pockets, if any, from the protein-membrane interface.

In summary, the three different regions that are readily available for further processing after the grid-labeling step are:

1. Solute region:   **insas**(i,j,k)>0
2. Membrane region: **insas**(i,j,k)<0 and **inmem**(i,j,k)>0
3. Solvent region:  **insas**(i,j,k)<0 and **inmem**(i,j,k)=0

**Membrane pore/channel detection**

Step 3 in the above general grid-labeling algorithm is meant to identify pore- or channel-like water-accessible water pockets within a user-specified membrane region. Given the convention that the membrane is parallel with the *xy* plane, the membrane region can be mathematically defined to be all grid points within [**zmin, zmax**]. Thus the method starts by initializing all grid points that are defined as solvent (**insas**<0) within [**zmin, zmax**] as **inmem**=1. Next the recursive depth-first search algorithm is used to traverse all grid points to see whether they are connected or not. Our goal of using the algorithm is to walk and label recursively all grid points in the non-protein regions within [**zmin, zmax**]. Upon completion, all grid points that are not connected to the membrane region (i.e. the pore region) are labeled back as the water region (**inmem**=0). To facilitate the bookkeeping of the search, the variable **kzone** is introduced to label the different regions: the protein region (**kzone**=0), the membrane region (**kzone**=1), and the water regions (**kzone**>1). Since the search starts from the edge of the membrane slab, the first region found is always the membrane region (**kzone**=1), and the rest are the water regions or the protein region. In general multiple **kzone** values are assigned because most water-accessible regions are not connected. The algorithm can be summarized as shown below:

```
nzone = 0; kzone = -1
for k = zmin:zmax
   for j,i = 1:n
      if kzone(i,j,k) != -1 then
         cycle
      end if
      if insas(i,j,k) > 0 then
         kzone(i,j,k) = 0
      else
         nzone = nzone + 1
         kzone(i,j,k) = nzone
         call walk(i,j,k,kzone,nzone)
      end if
   end
end

recursive subroutine walk(i,j,k,kzone,nzone)
   kzone(i,j,k) = nzone
   if (kzone(i+1,j,k) == -1 .and. insas(i+1,j,k)<0)
   call walk(i+1,j,k,kzone,nzone)
   if (kzone(i-1,j,k) == -1 .and. insas(i-1,j,k)<0)
   call walk(i-1,j,k,kzone,nzone)
   if (kzone(i,j+1,k) == -1 .and. insas(i,j+1,k)<0)
   call walk(i,j+1,k,kzone,nzone)
   if (kzone(i,j-1,k) == -1 .and. insas(i,j-1,k)<0)
   call walk(i,j-1,k,kzone,nzone)
   if (k+1<=zmax .and. kzone(i,j,k+1) == -1 .and. insas(i,j,k+1)<0)
   call walk(i,j,k+1,kzone,nzone)
   if (k-1>=zmin .and. kzone(i,j,k-1) == -1 .and. insas(i,j,k-1)<0)
   call walk(i,j,k-1,kzone,nzone)
end recursive subroutine walk
```

In this way, all grid points with **kzone**>1 are water accessible, and **inmem** of these grid points are set back to 0, i.e. membrane inaccessible.

**Mapping solvent/membrane accessibility to dielectric constants**

In this study, we adopted a three-dielectric model to model the membrane-protein electrostatics. The dielectric constants for the three different regions are denoted as $\varepsilon_{in}$ (solute), $\varepsilon_{out}$ (solvent) and $\varepsilon_{mem}$ (membrane), respectively.

The next step is to map the grid labeling information into the dielectric constants at the midpoints on all grid edges. The general principle, to be consistent with Wang et al, [43] is that the dielectric constant of a grid edge should be equal to the dielectric constant in the region where the two flanking grid points reside. When the two neighboring grid points belong to different dielectric regions, the weighted harmonic averaging (WHA) method is used to calculate the "fractional" dielectric constant based on the precise intersection point where the molecular surface cut the grid edge [43]. Specifically the dielectric constant is assigned as:

$$\varepsilon = \frac{1}{\frac{a}{\varepsilon_1} + \frac{1-a}{\varepsilon_2}}, \quad (4)$$

where $a$ denotes the fraction of the grid edge in region 1. Eqn (4) is applied on three different kinds of interfaces:

$$\begin{array}{ll} \varepsilon_1 = \varepsilon_{in}, \varepsilon_2 = \varepsilon_{out} & \text{solute and solvent interface} \\ \varepsilon_1 = \varepsilon_{in}, \varepsilon_2 = \varepsilon_{mem} & \text{solute and membrane interface} \\ \varepsilon_1 = \varepsilon_{out}, \varepsilon_2 = \varepsilon_{mem} & \text{solvent and membrane interface} \end{array} \quad (5)$$

The assignment of dielectric constants on the solute and solvent interface is the same as Wang et al [43]. We now consider the grid edge between (i,j,k) and (i+1,j,k); the grid edge can be classified according to the rules in Table I. The procedure of assigning the dielectric constants on the membrane related region and interface is as follows, for each of the x-, y-, and z-edges, respectively.

| insas(i,j,k) | insas(i+1,j,k) | inmem(i,j,k) | inmem(i+1,j,k) | region |
|---|---|---|---|---|
| >0 | >0 | >0 | >0 | inside solute |
| >0 | >0 | >0 | =0 | inside solute |
| >0 | >0 | =0 | >0 | inside solute |
| >0 | >0 | =0 | =0 | inside solute |
| >0 | <0 | >0 | >0 | solute and membrane interface |
| >0 | <0 | >0 | =0 | solute and solvent interface |
| >0 | <0 | =0 | >0 | solute and membrane interface |
| >0 | <0 | =0 | =0 | solute and solvent interface |
| <0 | >0 | >0 | >0 | solute and membrane interface |
| <0 | >0 | >0 | =0 | solute and membrane interface |
| <0 | >0 | =0 | >0 | solute and solvent interface |
| <0 | >0 | =0 | =0 | solute and solvent interface |
| <0 | <0 | >0 | >0 | inside membrane |
| <0 | <0 | >0 | =0 | inside solvent |
| <0 | <0 | =0 | >0 | inside solvent |
| <0 | <0 | =0 | =0 | inside solvent |

**Table I:** Different edges of dielectric constants defined by adjacent values of insas and inmem.

For x-edges, fractional membrane edges are only possible with the membrane-solute interface, so that the following pseudo code can be added to the existing dielectric mapping procedure:

```
If (inmem(i,j,k)>0 .or. inmem(i+1,j,k)>0) then
   If (insas(i,j,k)>0 .and. insas(i+1,j,k)>0) then
      εx(i,j,k) = εin
   // grid edge in solute
   else if ((insas(i,j,k)>0 .and. inmem(i+1,j,k)>0) .or.
   (insas(i+1,j,k)>0 .and. inmem(i,j,k)>0) ) then
```

$$\varepsilon_x(i,j,k) = \frac{1}{\dfrac{a}{\varepsilon_{in}} + \dfrac{1-a}{\varepsilon_{mem}}}$$

```
   // grid edge between membrane and solute
   else if (insas(i,j,k)>0 .or. insas(i+1,j,k)>0) then
```

$$\varepsilon_x(i,j,k) = \frac{1}{\dfrac{a}{\varepsilon_{in}} + \dfrac{1-a}{\varepsilon_{out}}}$$

```
   // grid edge between solvent and solute
   else if(inmem(i,j,k)>0 .and. inmem(i+1,j,k)>0) then
      εx(i,j,k) = εmem
   // grid edge in membrane
   end if
```

```
   end if
```

Here *a* is the fraction of grid edge in the solute region. The algorithm along the y-axis is similar to the x-axis, as follows:

```
If (inmem(i,j,k)>0 .or. inmem(i,j+1,k)>0) then
   If (insas(i,j,k)>0 .and. insas(i,j+1,k)>0) then
```
$$\varepsilon_y(i,j,k) = \varepsilon_{in}$$
```
   // grid edge in solute
   else if ((insas(i,j,k)>0 .and. inmem(i,j+1,k)>0) .or.
   (insas(i,j+1,k)>0 .and. inmem(i,j,k)>0) ) then
```
$$\varepsilon_y(i,j,k) = \frac{1}{\dfrac{a}{\varepsilon_{in}} + \dfrac{1-a}{\varepsilon_{mem}}}$$
```
   // grid edge between membrane and solute
   else if (insas(i,j,k)>0 .or. insas(i,j+1,k)>0) then
```
$$\varepsilon_y(i,j,k) = \frac{1}{\dfrac{a}{\varepsilon_{in}} + \dfrac{1-a}{\varepsilon_{out}}}$$
```
   // grid edge between solvent and solute
   else if(inmem(i,j,k)>0 .and. inmem(i,j+1,k)>0) then
```
$$\varepsilon_y(i,j,k) = \varepsilon_{mem}$$
```
   // grid edge in membrane
   end if
end if
```

For the dielectric constant mapping along the z-axis, the algorithm also involves the solvent-membrane interface; the algorithm should also take care of this, as follows:

```
If (inmem(i,j,k)>0 .or. inmem(i,j,k+1)>0)
   If (insas(i,j,k)>0 .and. insas(i,j,k+1)>0) then
```
$$\varepsilon_z(i,j,k) = \varepsilon_{in}$$
```
   else if ((insas(i,j,k)>0 .and. inmem(i,j,k+1)>0) .or.
   (insas(i,j,k+1)>0 .and. inmem(i,j,k)>0) ) then
```
$$\varepsilon_z(i,j,k) = \frac{1}{\dfrac{a}{\varepsilon_{in}} + \dfrac{1-a}{\varepsilon_{mem}}}$$
```
   // grid edge between membrane and solute
   else if (insas(i,j,k)>0 .or. insas(i,j,k+1)>0) then
```
$$\varepsilon_z(i,j,k) = \frac{1}{\dfrac{a}{\varepsilon_{in}} + \dfrac{1-a}{\varepsilon_{out}}}$$
```
   // grid edge between solvent and solute
   else if(inmem(i,j,k)>0 .and. inmem(i,j,k+1)>0) then
```

$$\varepsilon_z(i,j,k) = \varepsilon_{mem}$$
```
    else if (grid edge is cross the slab)
```
$$\varepsilon_z(i,j,k) = \frac{1}{\dfrac{a}{\varepsilon_{out}} + \dfrac{1-a}{\varepsilon_{mem}}}$$
```
    // grid edge cross the slab
    // a is fraction of edge in solvent
    end if
  end if
```

Finally, all the edges in the water are assigned the dielectric constant of water, all the edges in the membrane are assigned the dielectric constant of membrane, and all the edges in the protein interior are assigned the dielectric constant of protein. For all the edges crossing different regions, i.e. between any two of water, membrane, or protein, weighted harmonic averages between the two corresponding dielectric constants are assigned.

**Protein and complex structure preparation**

To calibrate the new continuum membrane model for channel detection, we simulated three channel proteins with crystal structures: 1K4C [136], a KcsA potassium channel; 5CFB [137], an alpha1 GlyR Glycine receptor; and 5HCJ [138], a prokaryotic pentameric ligand-gated ion channel. To demonstrate the feasibility of the new continuum membrane model in data intensive binding affinity calculations, we chose the hERG K+ channel protein, given its importance in drug discovery and the availability of high-quality experimental data [139].

A homology model of hERG K+ channel was built based on the X-ray crystal structure of KcsA [136] (PDB ID: 1K4C) using MODELLER [140] (version 9.15) with the default settings. The amino acid sequence of hERG K+ channel was directly extracted from the Swiss-Prot database [141] (accession number: Q12809 and entry: KCNH2_HUMAN). Sequence alignment was generated using CLUSTALX [142] (version 2.1), showing a good match in helices S5, S6, and the pore region, with identity about 44% (Figure 1). After automatic model building and loop refinement, candidate models were evaluated based on the DOPE score from MODELLER [140]. The final homology model of the hERG K$^+$ channel is shown in Figure 2, which is found to be highly consistent with a previously reported model based on a different procedure [143-146].

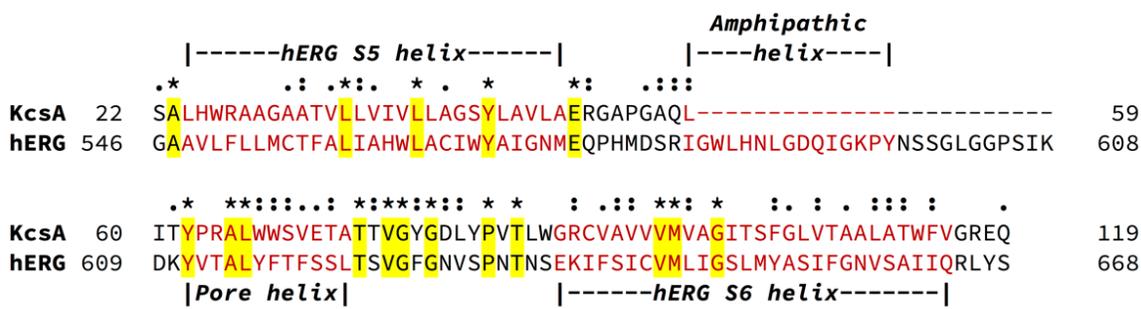

**Figure 1**: Sequence alignment of KcsA and hERG by ClustalX version 2.1. The identified S5 helix, S6 helix, amphipathic helix and pore helix are labeled above the sequence. Asterisks (*): conserved amino acid residues; colons (:): conserved substitutions; dots (.): semi-conserved substitutions.

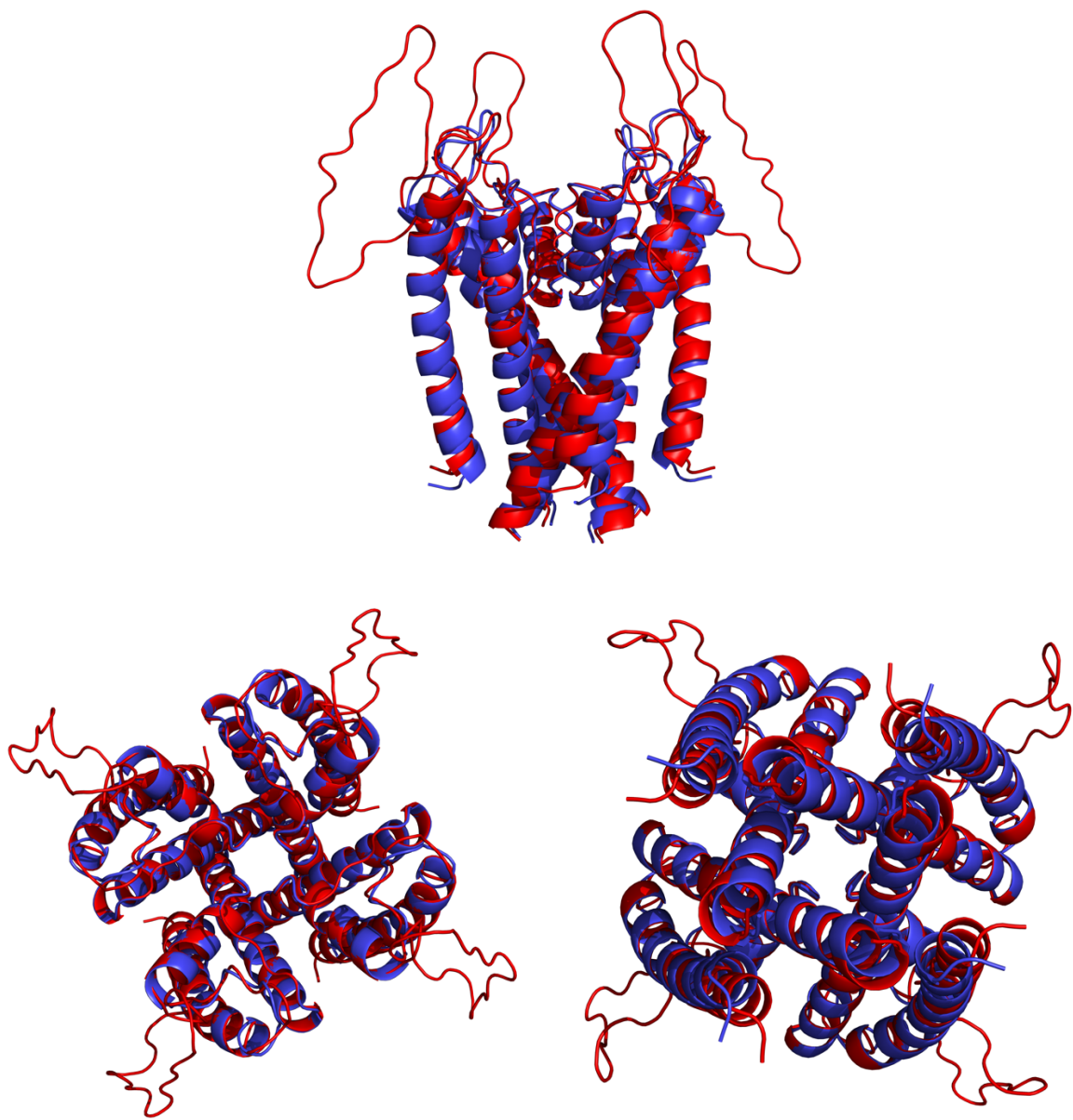

**Figure 2**: Comparison of target and parent structures, showing the secondary structure elements in homology models of hEGH (red) and KcsA (blue). The plot shows three orientations of the aligned structure. Top: side view with the binding pocket on the top. Bottom left: viewed from the binding pocket/extracellular side. Bottom right: viewed from the intracellular side.

Initial complex structures of the hERG K+ channel with its inhibitors were generated with the SURFLEX-DOCK program in Sybyl-X (version 1.3). Ten different inhibitors with experimental binding affinities [139] were chosen to assess the quality of the MMPBSA procedure, including astemizole (AST), sertindole (SER), pimozide (PIM), droperidol (DRO), terfenadine (TE0, TE1), domperidone (DOM), loratadine (LOR), mizolaatine (MIZ), perhexiline (PE0, PE1) and amitriptyline (AMI). The terfenadine and perhexiline are chiral molecules with two enantiomers, so both enantiomers were used in the docking.

**Molecular dynamics simulation**

The protein was first inserted into a membrane layer using the CHARMM-GUI lipid builder [147-151]. Lipid DPPC was used for the membrane layer with a lipid to water ratio of 29. The solvated membrane system first underwent a 10,000-step energy minimization using a 5,000-step steepest descent followed by a 5,000-step conjugate gradient. The main chain atoms for the protein were then restrained with a force constant of 2 kcal/mol-$Å^2$. Subsequently, a 5 ps MD simulation was conducted to heat the system from 0 to 100K followed by a 100 ps MD simulation to heat the system from 100K to 310K. This was then followed with a 5 ns simulation for equilibration. Finally, production MD was run for 50 ns.

**MMPBSA calculations of binding affinities**

Binding free energies were computed using a revised MMPBSA module[124] of Amber 16 or AmberTools 2016 [134, 135, 152]. The production run trajectory was post-processed with CPPTRAJ [153] in order to remove the solvent, membrane, and counter ions from the receptor-ligand complex. Snapshots from the last 10 ns of the production run were processed to compute molecular mechanics potential energies and solvation free energies in the MMPBSA procedure. The binding free energy for the protein-ligand complex was computed as the difference between the complex free energy and the sum of the receptor and ligand free energies, as outlined in our previous work [124]. The electrostatic solvation free energies were calculated using the linearized PBE model as implemented in PBSA [36, 43, 80, 82, 87]. The non-electrostatic solvation free energies were calculated using either the classical model or the modern model as documented previously [154].

**Additional computational details**

In each PBSA calculation, a finite-difference grid spacing of 0.5 Å was used for MMPBSA calculations, which was found to be sufficient due to MD sampling and the approximate nature of the binding affinity calculation [118]. Production snapshots up to 10ns were found to be sufficient to converge the averaging process used in MMPBSA calculations of these membrane protein-ligand complexes. The periodic geometric multigrid solver option was employed with a convergence threshold of $1.0 \times 10^{-3}$, and electrostatic focusing was turned off due to the presence of the membrane [87]. The use of a periodic boundary also allowed a somewhat small fillratio (i.e. the ratio of the finite-difference box dimension over the solute dimension) of 1.25 to be used in these calculations [37]. The solvation system physical constants were set up as follows. The membrane was modeled as a solid slab as simulated in the explicit water MD trajectories. The water relative dielectric constant was set at 80.0. The membrane dielectric constant was set to be 7.0 [124]. And the protein dielectric constant was set to be 20.0 due to the presence of charged ligand molecules [118, 124]. The water phase ionic strength was set to be 150 mM. The lower dielectric region within the molecular solutes was defined with the classical solvent excluded surface model using a water solvent probe and a membrane solvent probe to be optimized as described in Results and Discussion. The default weighted harmonic averaging was employed to assign dielectric constants for boundary grid edges to reduce grid dependency [43]. Charges and radii were assigned as in the simulation topology files.

## Results and Discussion

**Optimization of the new slab membrane model**

Given the automatic procedure in place to identify water channels/pores with the depth-first search method, we further optimized the membrane probe value and the slab membrane model (i.e. its thickness) to best reproduce the distributions of buried water molecules in the membrane region as sampled in explicit water MD simulations. Three different membrane proteins with channels were utilized in this optimization: 1K4C, 5HCJ, and 5CFB.

Three different slab definitions were evaluated to set up the continuum membrane model, i.e. the inner and outer faces are chosen to be positioned at (1) the average z-coordinates of nitrogen atoms of the lipid head groups; (2) the average z- coordinates of the phosphorus atoms of the lipid head groups; (3) the average z- coordinates of both nitrogen and phosphorus atoms in the lipid head groups. Here the average z-coordinates are computed from the explicit-water MD simulations.

Next, `mprob` values were scanned from 1.4 Å upwards to 3.0 Å with an increment of 0.1 Å. The smallest `mprob` value with which these known channels can be displayed was recorded as the `mprob` threshold in Table II for all three slab membrane definitions. It should be pointed out that a small `mprob` produces excessive membrane accessibility in the protein interior so that it is more likely for the buried membrane pockets to be connected to the bulk membrane. Excessive membrane accessibility can also be lessened by reducing the membrane thickness, as in the use of phosphorus atoms to define the boundaries of the continuum membrane. Indeed, our analysis showed this setup caused the least penetration of the continuum membrane into the protein interior, so the smallest mprob (2.7 Å) was needed to capture the water channels/pores for all three tested proteins.

Figure 3 shows the rendering of water-channels/pores of the three tested membrane proteins with the optimized `mprob`. The advantage of the optimal mprob over the default solvent probe of 1.4 Å is apparent by comparing the renderings generated with the two probes. For all the channel proteins, the new model automatically detects the water channels/pores. Figure 4 further shows the benefit of the depth-first-search feature that is a must in the new slab membrane model. Without it, it is apparent that none of the water channels/pores can be identified for any of the tested proteins, even using the larger probe for the membrane region.

| Protein | mthick (Å) | mcenter (Å) | mprob (Å) |
|---|---|---|---|
| | mthick=$|N^+–N^-|$ | | |
| 1K4C | 39.24 | -1.10 | >2.2 |
| 5CFB | 40.72 | 64.95 | >1.7 |
| 5HCJ | 41.11 | 68.60 | >3.0 |
| | mthick=$|P^+–P^-|$ | | |
| 1K4C | 36.13 | -0.97 | >2.2 |
| 5CFB | 37.27 | 64.97 | >1.6 |
| 5HCJ | 37.89 | 68.40 | >2.7 |
| | mthick=$|N^+P^+– N^-P^-|$ | | |
| 1K4C | 37.69 | -1.04 | >2.2 |
| 5CFB | 39.00 | 64.96 | >1.7 |
| 5HCJ | 39.50 | 68.50 | >3.0 |

**Table II** The thickness of membrane and mprob thresholds based on the different criterion measured from MD simulations. Here the mprob threshold is the minimum value with which the channel is visible with the SES approach. (Top) mthick=$|N^+–N^-|$: The thickness of the membrane slab is defined as the z-distance between the average head group nitrogen atoms of the lipid molecules. (Middle) mthick=$|P^+–P^-|$: The thickness of the membrane slab is defined as the z-

distance between the average head group phosphorus atoms of the lipid molecules. (Bottom) mthick=$|N^+P^+- N^-P^-|$: the membrane slab is defined as the z-distance between the average head group centers (i.e. the means of nitrogen and phosphorus atoms) of the lipid molecules. The membrane center locations were then computed as the mean of the upper and lower bounds.

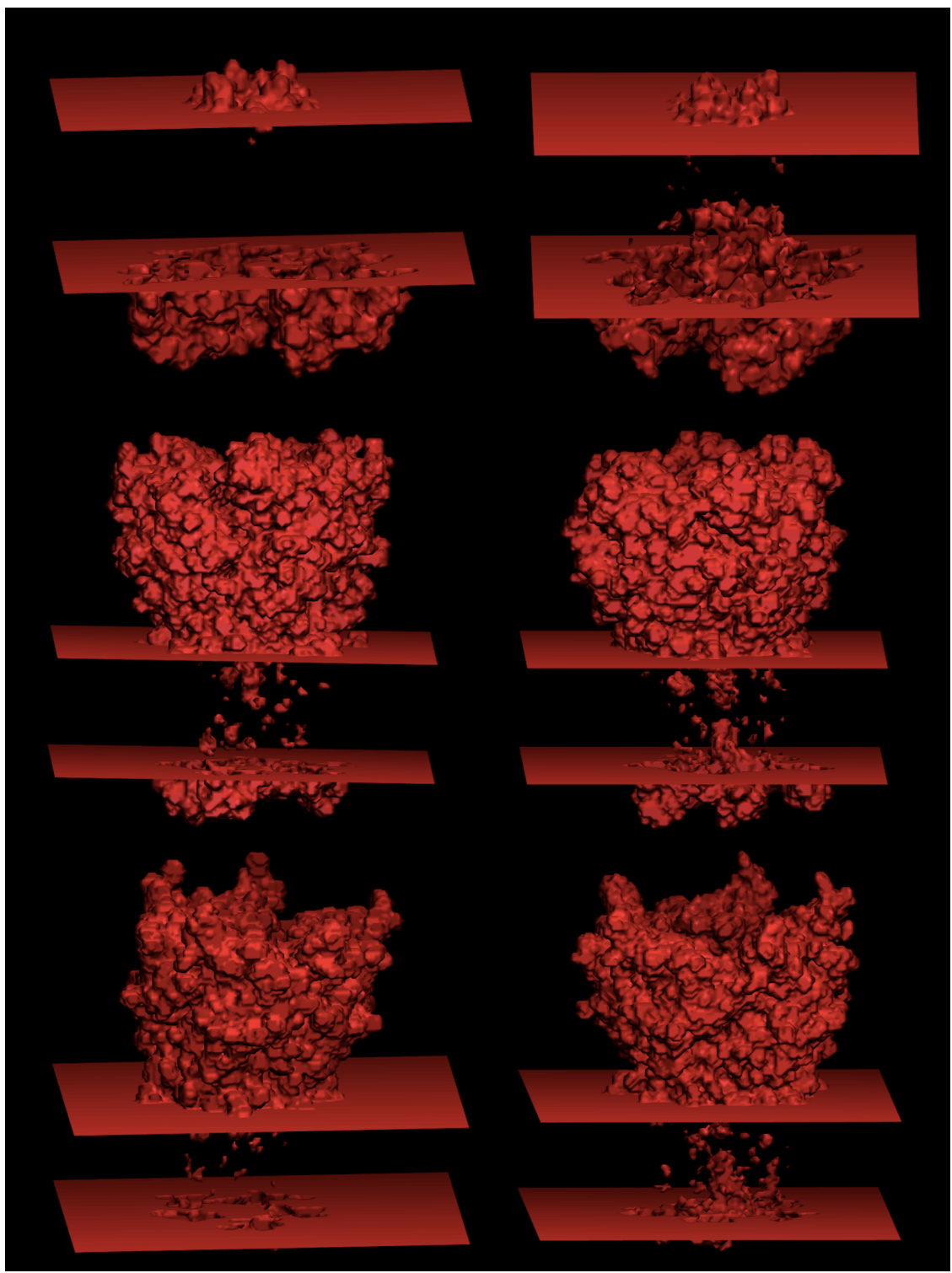

**Figure 3** Solvent-solute interface determined with the new continuum membrane model. Left: mprob is set to be 1.4 Angstroms, the default value of the solvent probe. Right: mprob is set to be 2.7 Angstroms, the optimized value of the membrane probe. Three proteins are tested: 1K4C (top); 5CFB (middle); 5HCJ (bottom).

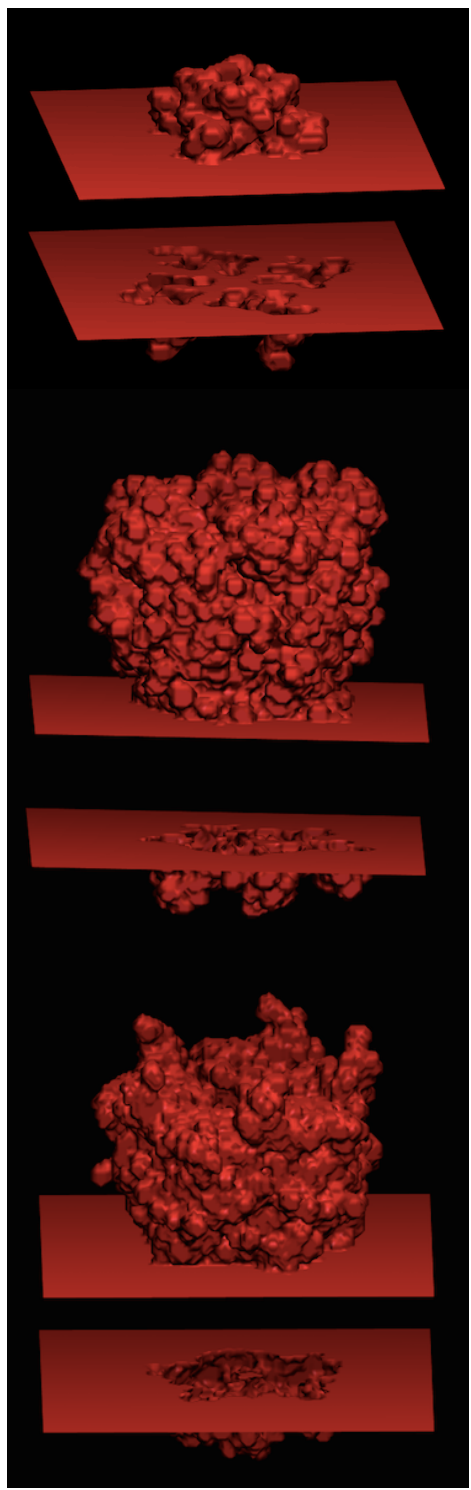

**Figure 4** Same as Figure 3, except without turning on the depth-first search in the pore region detection. Three proteins are tested: 1K4C (top); 5CFB (middle); 5HCJ (bottom).

**Impact of the water solvent probe upon agreement with an explicit solvent simulation**

It is worth pointing out that the agreement of the continuum membrane model also depends on how we model the water accessible region. The standard practice has been to consider the finite size of the water molecule with a predefined probe radius, often taken as 1.4 Å. The probe is then used to compute the solvent excluded surface used as the interface separating the protein interior from the water region. It is apparent that the size of the water accessible pores/channels would depend on how large the water probe is defined. Thus, it is interesting to analyze how well the widely used water probe performs in the context of membrane channel proteins.

This analysis was conducted in the following manner. The distributions of water molecules (in the water pore/channel regions) in explicit water MD simulations were sampled every 50 ps over the course of a 5 ns production run. Note that the protein atoms were all restrained to the reference structure after equilibration since the focus was on the water distribution. A total of 100 frames worth of water sampling were collected and were combined into one snapshot for visualization. This water distribution map was used as a reference to evaluate how the hard sphere SES surface behaves with one single adjustable parameter, i.e. the water solvent probe (`dprob` in Amber/PBSA).

The same three membrane channel proteins were analyzed to address this question. Specifically, the counts for the following disagreements/mismatches were recorded: (1) the absence of explicit water molecules in the continuum water accessible regions; and (2) the presence of explicit water molecules in the continuum water inaccessible regions. The overall summary of both mismatches is reported in Table III. Sample mismatches are shown in Figure 5. It is interesting to note that the standard value of the water solvent probe of 1.4 Å is a very reasonable default value, which gives an overall minimum number of inconsistencies between the continuum and explicit representations of water distributions in the tested membrane channel proteins, at least in the water accessible pore/channel regions that we have focused on.

| Protein | dprob (Å) | No. solvent region w/o water molecules | No. water molecules in non-solvent region |
| --- | --- | --- | --- |
| 1K4C | 1.2 | 23 | 4 |
| | 1.3 | 18 | 8 |
| | 1.4 | 15 | 11 |
| | 1.5 | 14 | 14 |
| | 1.6 | 12 | 15 |
| 5CFB | 1.2 | 20 | 5 |
| | 1.3 | 17 | 10 |
| | 1.4 | 14 | 11 |
| | 1.5 | 10 | 17 |
| | 1.6 | 7 | 22 |
| 5HCJ | 1.2 | 22 | 4 |
| | 1.3 | 18 | 7 |
| | 1.4 | 9 | 11 |
| | 1.5 | 9 | 17 |
| | 1.6 | 7 | 23 |

**Table III** Discrepancies in the solvent accessible region between explicit water MD simulations and the membrane PBSA calculations. Two types of discrepancies were recorded: (1) how many continuum solvent pockets do not have water molecules; and (2) how many explicit water molecules are observed in the non-continuum solvent pockets defined in the membrane PBSA calculation. The water probe (dprob) was scanned from 1.2 Å to 1.6 Å for three different proteins: 1K4C, 5CFB, 5HCJ. The membrane setup has been optimized according to the values given in Table II. For each protein, the samples of water molecules were taken from a 5ns equilibrium MD simulation with all protein atoms restrained to the initial structure, which was obtained from the last snapshot of the unconstrained normal MD, which is also the reference for the water sampling run. The listed values are the averages of 100 snapshots evenly selected from the 5ns MD simulation.

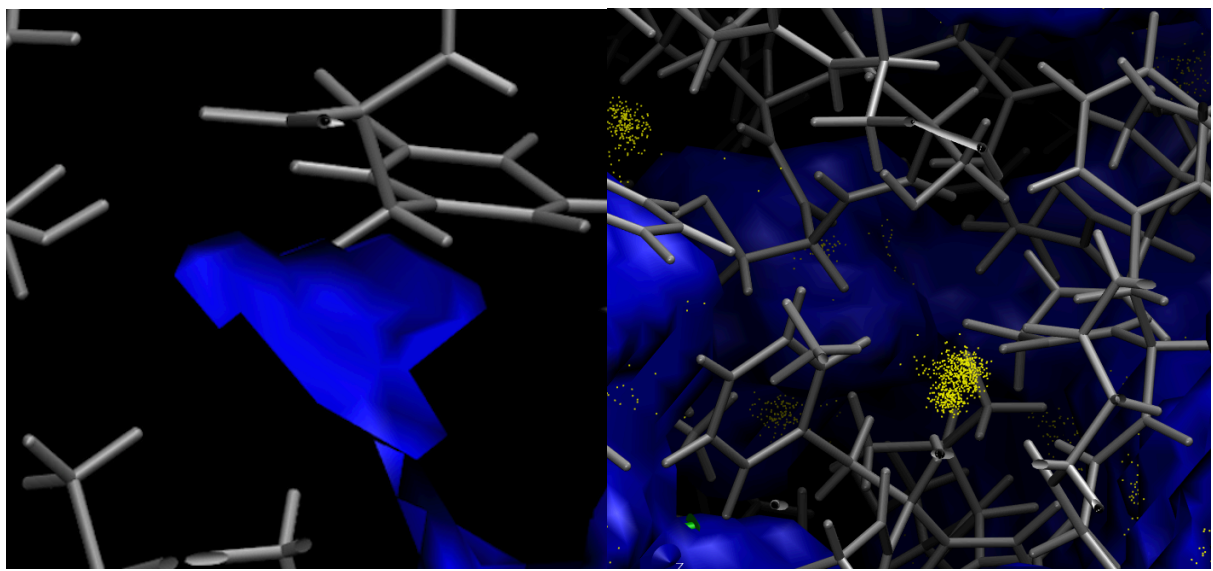

**Figure 5** Discrepancy between implicit and explicit water simulations. The protein surface of 1K4C (blue) is overlaid with a bond representation and sampled water positions (yellow). Left: a solvent region defined by the PBSA model but with no explicit water. Right: explicit water is detected in a region where no solvent is defined in the PBSA model.

It is instructive to point out that the inconsistency between the two representations may be due to the setup of the explicit water MD simulation and also to the limitations of MD sampling of water distributions. First, it is well known that isolated water cavities exist in the protein interior, which are disconnected from the bulk water. Unless crystal water molecules were observed and retained in the initial setup of the MD simulations, these isolated cavities are most likely modeled as water-free due to the default closeness tolerance used in the placement of explicit water molecules when building the topology files. This issue would lead to the type (1) mismatches described above.

Second, although protein atoms were restrained during the MD simulations, they are not as inflexible as frozen hard spheres as in the case of the continuum solvent model that must use a single mean structure as input. Their motions allow minor structural changes, leading to the opening and closing of buried water cavities. If the mean structure happens to correspond to a closed form, the continuum model would not capture the water-accessible cavity.

Finally, the protein atom cavity radii that were used to present the size of each atom were chosen to be best for energetics and/or stability of the MD simulations. These may or may not be optimal to quantify water accessibility in the protein interior. This points to future efforts to model the protein-water interface more self-consistently based on the consistent energy model as defined by the protein-water force field used in both explicit and implicit simulations.

**MMPBSA calculations of binding affinities**

Finally, as an illustration of our new continuum membrane model, we conducted a set of binding free energy calculations of ten different ligands independently bound to a potassium channel protein. The computed binding affinities and experimental $IC_{50}$ values are summarized in Table IV. The correlation analysis between computation and experiment is shown in Figure 6. Both the classical and modern nonpolar solvent models (INP=1 and INP=2 respectively) were tested, and the correlations for these two methods are similar, which is consistent with what we expected. Overall good correlations with experiment were observed: with correlation coefficients of 0.79 for INP=1 and 0.73 for INP=2 (due to the smaller range of the data).

| Name | RTln(IC50) | mthick | mcenter | MMPBSA (INP=1) | MMPBSA (INP=2) |
|---|---|---|---|---|---|
| Amitriptyline | -7.08 | 36.086 | -10.383 | -38.23 | -9.84 |
| Perhexline (PE0) | -7.23 | 36.661 | -1.620 | -40.73 | -12.44 |
| Perhexline (PE1) | -7.23 | 36.473 | -5.681 | -41.96 | -13.63 |
| Mizolastine | -9.15 | 36.260 | 1.360 | -61.25 | -24.59 |
| Loratadine | -9.58 | 36.126 | -0.969 | -52.48 | -18.64 |
| Domperidone | -9.62 | 35.979 | 0.707 | -59.53 | -20.69 |
| Terfenadine (TE0) | -9.82 | 36.616 | -2.319 | -69.67 | -29.00 |
| Terfenadine (TE1) | -9.82 | 37.216 | -0.611 | -72.91 | -30.56 |
| droperidol | -10.61 | 36.459 | 3.308 | -59.04 | -24.78 |
| Pimozide | -10.97 | 36.910 | 0.190 | -60.04 | -20.09 |
| Sertindole | -11.13 | 36.438 | -1.161 | -62.91 | -23.89 |
| Astemizole | -12.81 | 34.133 | -0.043 | -67.64 | -26.21 |

**Table IV:** MMPBSA binding affinities (kcal/mol) in comparison with experiment (IC50). The slab membrane geometry (the thickness and z-center in Å) compiled from the explicit solvent MD simulation are also shown for each complex.

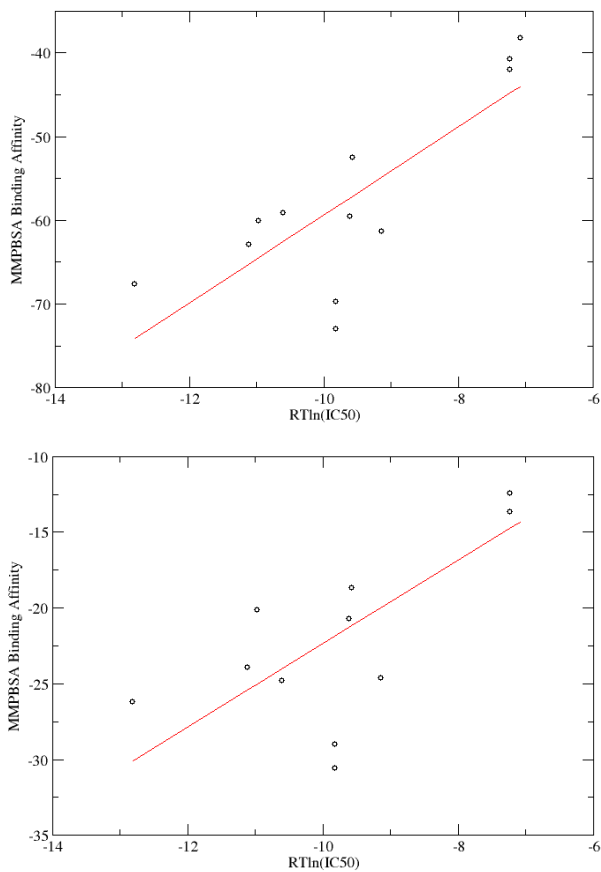

**Figure 6:** MMPBSA binding affinities compared with experimental measurements. Binding affinities are in kcal/mol. Top: MMPBSA was computed with the classical nonpolar solvent

model (INP=1), the correlation coefficient is 0.79. Bottom: MMPBSA was computed with the modern nonpolar solvent model (INP=2), the correlation coefficient is 0.73.

**Timing analysis**

Finally, we conducted a timing analysis of the new membrane model. Table V summarizes the average CPU times over 100 frames that are used for setting up the dielectric grids with or without the membrane model in the MMPBSA calculation of the receptor. We can see the average time for the surface calculation increases by more than four times; this is mainly because two separate SES calls are made, once with the water probe and once with the membrane probe. Furthermore, the SES calculation with the much larger membrane probe is behind the much higher cost in the total SES time due to the longer non-bonded list and many more overlaps among larger probe-augmented atomic volumes. In addition, the grid-labeling step is also about three times slower, though not a significant portion of the overall CPU cost. Finally, the mapping from grid labels to dielectric constants changes little due to the virtually linear nature of the algorithm [43]. Overall the PBSA calculations are about 25% slower with the new continuum membrane model than those without any continuum membrane (i.e. modeled as a globular protein) for the tested protein-ligand binding calculations.

|  | Globular Protein Setup | Membrane Protein Setup |
|---|---|---|
| SES Calculations (s) | 3.76 | 15.63 |
| Grid Labeling (s) | 1.61 | 4.79 |
| EPS Mapping (s) | 0.21 | 0.22 |

**Table V:** Average CPU times (in seconds) used in setting up the dielectric grid for 100 snapshots in the MMPBSA calculation of the receptor. The membrane-free set up was run using memopt=0, and the membrane setup was run using memopt=1 in Amber/PBSA.

## Conclusions

We have proposed a new continuum membrane model for Poisson-Boltzmann calculations of biomolecules. Major improvements from the standard continuum slab model are the following: 1) explicit-solvent MD simulations were utilized to fine tune the slab model, i.e. its exact location and thickness, to best reproduce the solvent accessibility and the water accessible channel; 2) A two-step, two-probe initial grid labeling procedure was adopted to address highly different accessibility in the membrane region and water region; and 3) A depth-first search algorithm was introduced to detect the water pores/channels automatically based on the initial grid labels. This procedure follows our basic algorithm proposed for globular proteins and does not add significant overhead to the numerical PB calculations.

Given the revisions proposed above, we optimized the membrane probe value and the slab membrane model (i.e. its thickness) to best reproduce the distributions of buried water molecules in the membrane region as sampled in explicit water MD simulations. Three different membrane proteins with channels were utilized in this optimization. Our analysis showed that a slab membrane model using the mean phosphate atom positions as the membrane boundary and the smallest membrane probe of 2.7 Å caused the least penetration of the continuum membrane into the protein interior.

Apparently, the solvent accessibility also depends on how the continuum water is modeled. Thus, we used a water distribution map from an explicit water MD simulation as benchmark data to evaluate how the hard sphere SES behaves with one single adjustable parameter, i.e. the water solvent probe. The same three membrane channel proteins were analyzed to address this question. It is interesting to note that the standard value for the water solvent probe of 1.4 Å is very reasonable, which gives an overall minimum number of inconsistencies between the continuum and explicit representations of water distributions in the membrane channel proteins, at least in the water accessible pore/channel regions that we have focused on.

Finally, we conducted a set of binding affinity calculations of ten different ligands independently bound to a potassium channel using the new continuum membrane model. Both the classical and modern nonpolar solvent models were tested, and the correlations with experiment are similar with both models, which is consistent with our findings in globular proteins. Overall good correlations with experiment were observed, with correlation coefficients of 0.79 for INP=1 and 0.73 for INP=2. Finally, our timing analysis showed that the average time for the surface calculation increased by more than four times. The grid-labeling step is also about three times slower even though it is not a significant portion of the overall CPU cost. The mapping from grid labels to dielectric constants changed little due to the virtually linear nature of the algorithm.

Future efforts will be conducted to model the protein-water interface more self-consistently based on the consistent energy model as defined using the protein-water force field in both explicit and implicit simulations.

## Acknowledgements

This work is supported in part by the NIH (GM093040 & GM079383).